\documentclass{moriond}

\pdfoutput=1
\usepackage{graphicx} 
\usepackage{epstopdf} 
\usepackage{subfigure}
\usepackage{hyperref}
\usepackage{subfigure}
\usepackage{epsfig}
\usepackage{lineno}

\hypersetup{citecolor=blue,linkcolor=blue,linktoc=page,urlcolor=blue,colorlinks=true}

\def\be{\begin{equation}}
\def\ee{\end{equation}}
\def\bea{\begin{eqnarray}}
\def\eea{\end{eqnarray}}

\newcommand{\Photo}{} 

\begin{document}
\vspace*{0.7cm}
\title{Recent constraints on the parton distributions 
in the proton and the measurement of $\alpha_S$ from ATLAS and CMS}

\author{ M. R. Sutton} 

\address{Department of Physics and Astronomy, University of Sussex, 
Falmer, \\
Brighton BN1 9QH, United Kingdom}
\author{ On behalf of the \\ATLAS and CMS Collaborations }

\maketitle\abstracts{
\ \\
\ \\
Recent results on cross sections sensitive to the parton distribution 
functions (PDFs) within the proton from the ATLAS and CMS Collaborations 
are presented.  The potential impact on the inclusion of these data in 
fits to the PDFs is discussed. Recent results from fits including the
data from jet, or vector boson production from the ATLAS and CMS experiments 
are discussed.  
}

\newcommand{\suttfigure}[4]{
\begin{figure}[tph]
\centerline{\includegraphics[width=#2\linewidth]{#1}}
\caption{#3}
\label{#4}
\end{figure}
}

\newcommand{\suttdoublefigure}[6]{
\begin{figure}[tph]
\subfigure{\includegraphics[width=#2\linewidth]{#1}}
\subfigure{\includegraphics[width=#4\linewidth]{#3}}
\caption{#5}
\label{#6}
\end{figure}
}

\newcommand{\suttdoublescalefigure}[8]{
\begin{figure}[tph]
\subfigure{\includegraphics[width=#2\linewidth,height=#3\linewidth]{#1}}
\subfigure{\includegraphics[width=#5\linewidth,height=#6\linewidth]{#4}}
\caption{#7}
\label{#8}
\end{figure}
}

\newcommand{\figone}{
\suttdoublefigure{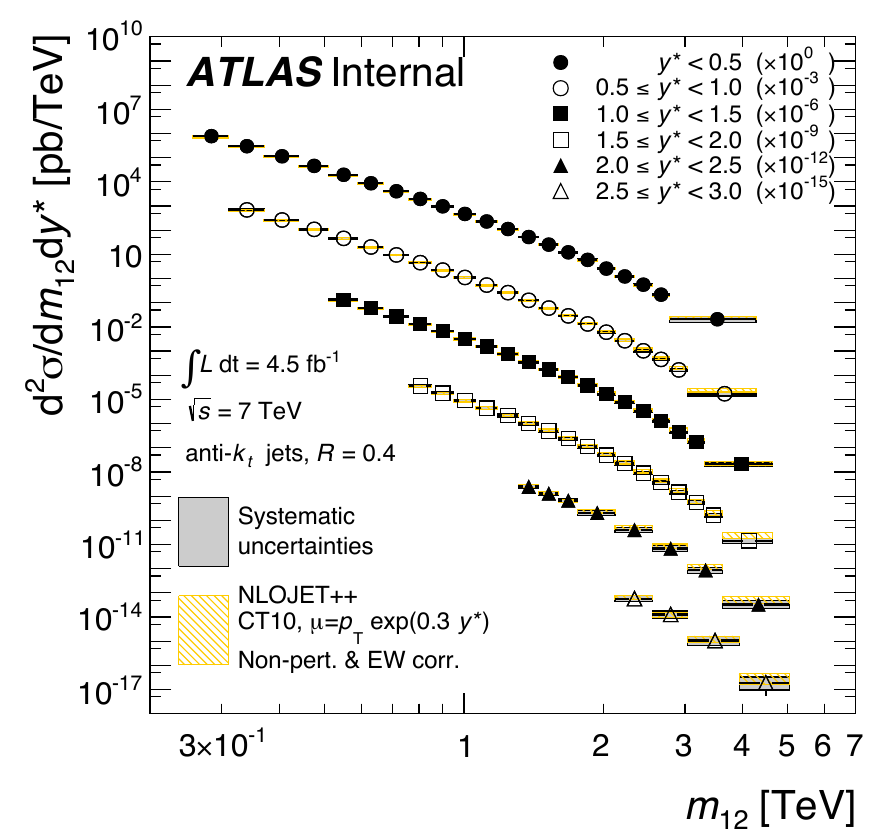}{0.49}{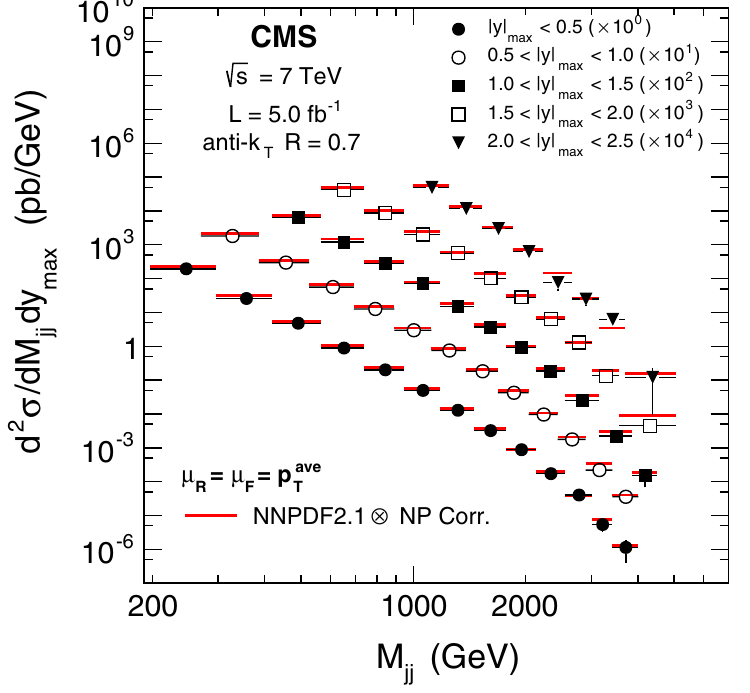}{0.49}
{The ATLAS (left) and CMS (right) dijet mass from the 2011 dijet data}{fig:1}
}

\newcommand{\figtwo}{
\suttfigure{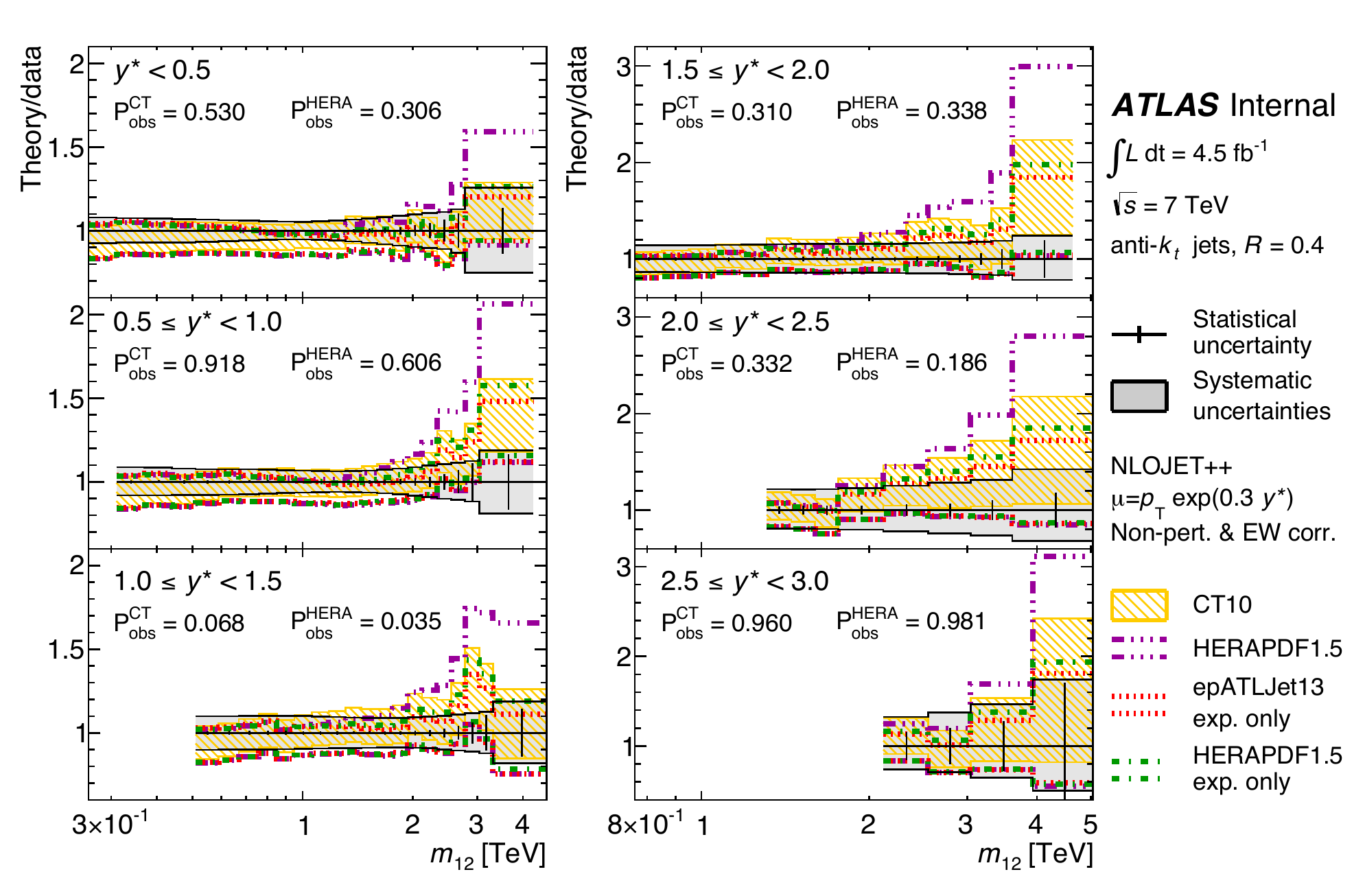}{0.9}{The ratio of the theory prediction to the ATLAS jijet data for different PDFs.}{fig:2}
}

\newcommand{\figthree}{
\begin{figure}[tph]
\subfigure{\includegraphics[width=0.49\linewidth]{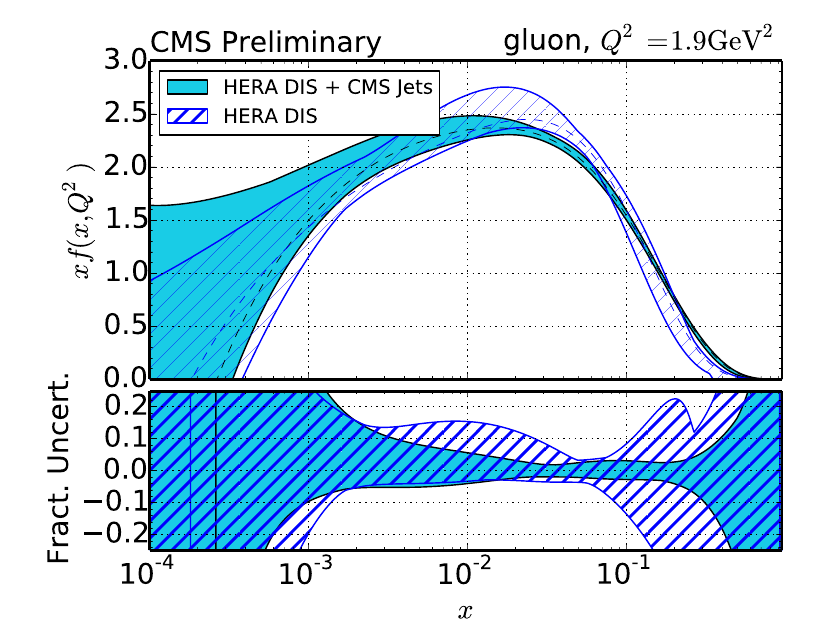}}
\subfigure{\includegraphics[width=0.486\linewidth]{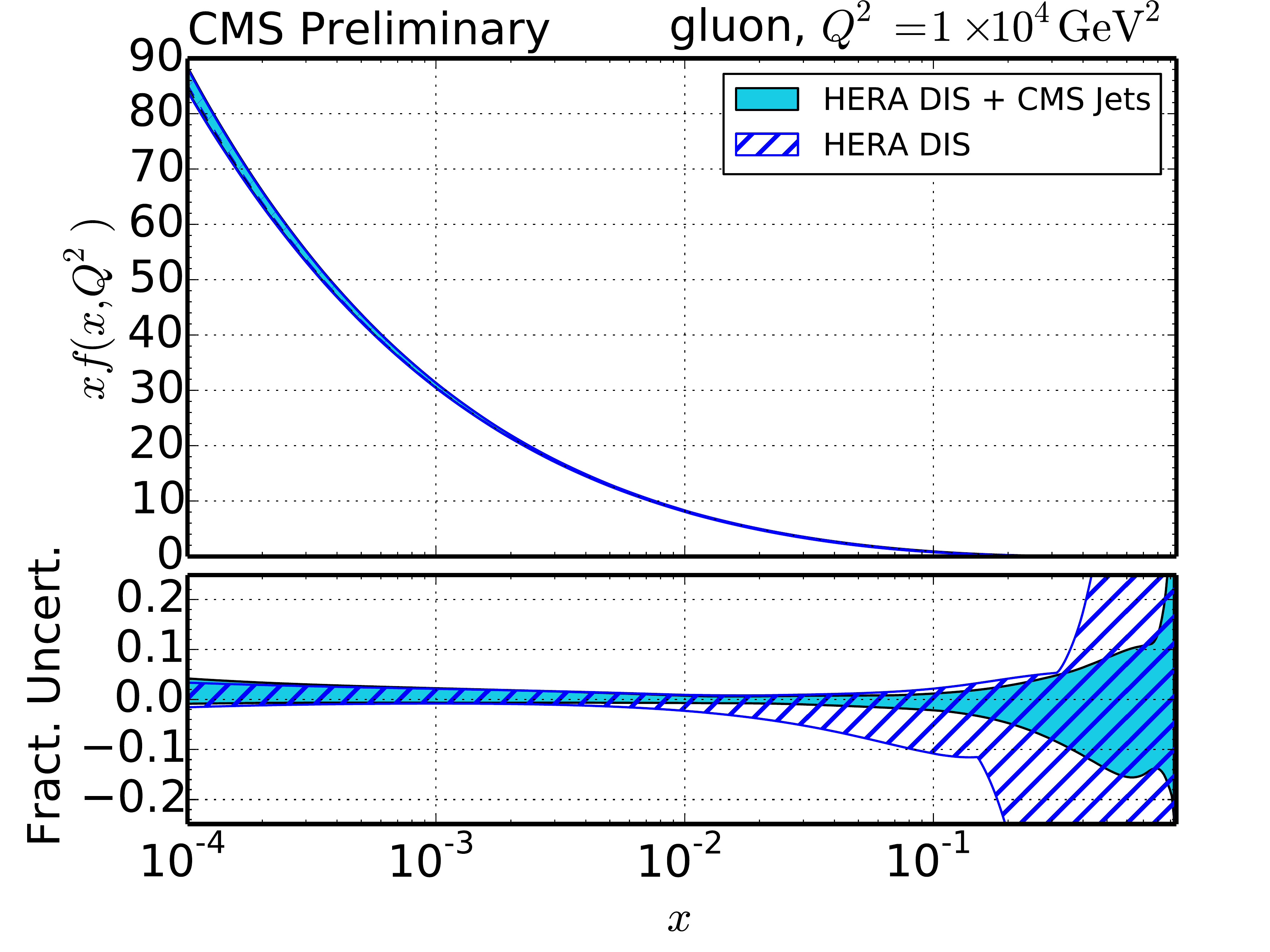}}
\vspace{-4mm}
\caption{The gluon distribution from the CMS fit; 
(left) at the starting scale of 1.9\,GeV$^2$ and (right) at a scale of 10$^4$\,GeV$^2$}
\label{fig:3}
\end{figure}
}

\newcommand{\figthreea}{
\begin{figure}[htp]
\begin{minipage}{0.63\textwidth}
\includegraphics[width=\textwidth]{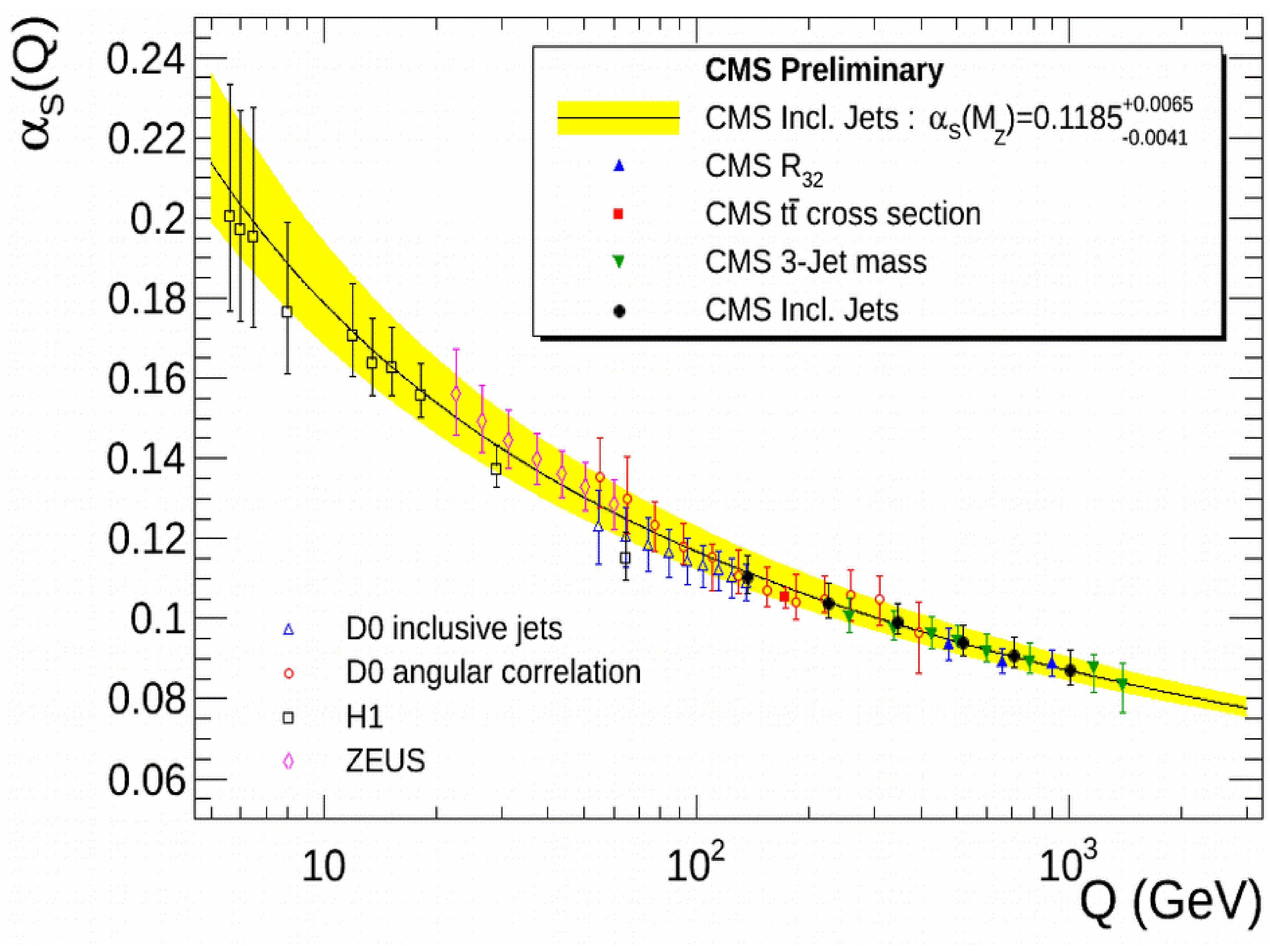}
\end{minipage}
\hspace{0.7cm}
\begin{minipage}{0.3\textwidth}\
\vspace{3cm}
\caption{
The strong coupling measured using the CMS data compared to measurements from HERA and the Tevatron.}
\label{fig:3a}
\end{minipage}
\end{figure}
}

\newcommand{\figfour}{
\begin{figure}[htp]
\begin{minipage}{0.6\textwidth}
\includegraphics[width=\textwidth]{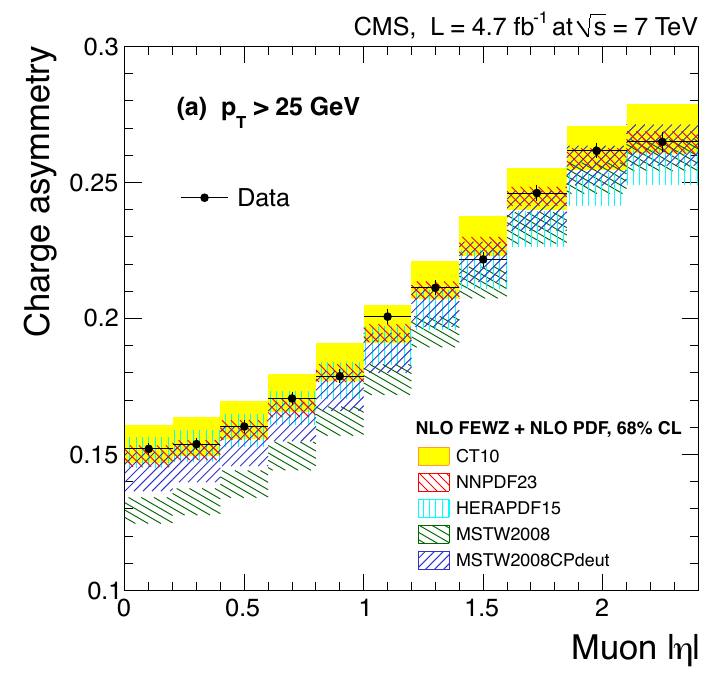}
\end{minipage}
\hspace{0.7cm}
\begin{minipage}{0.3\textwidth}\
\vspace{3.5cm}
\caption{
The CMS Electroweak charged $W$ asymmetry data from 2011.}
\label{fig:4}
\end{minipage}
\end{figure}
}

\newcommand{\figfive}{
\begin{figure}[thp]
\begin{minipage}{0.54\textwidth}
\includegraphics[width=\textwidth,height=0.7\linewidth]{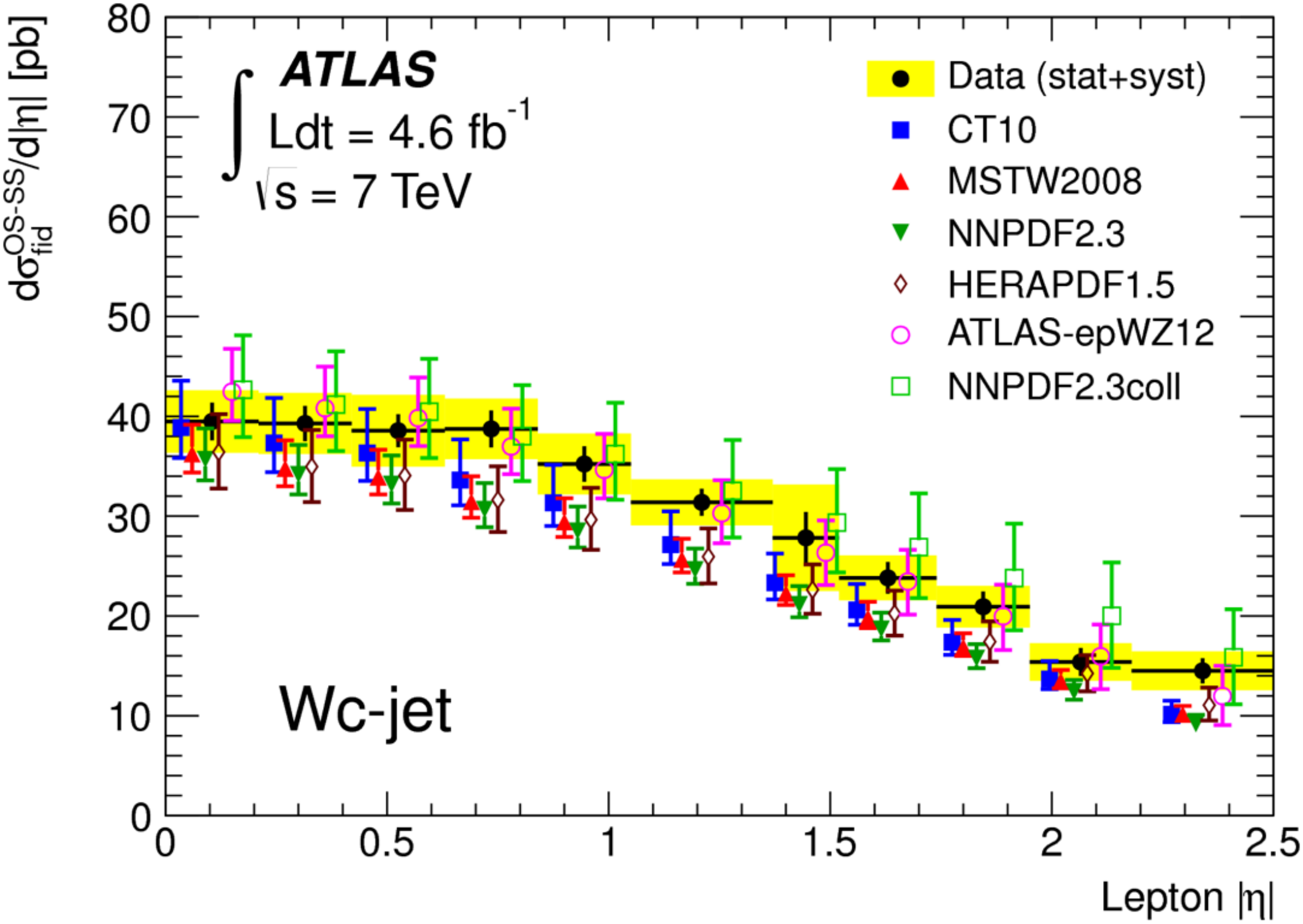}
\end{minipage}
\hspace{0.02\textwidth}
\begin{minipage}{0.44\textwidth}
\includegraphics[width=\textwidth,height=0.95\linewidth]{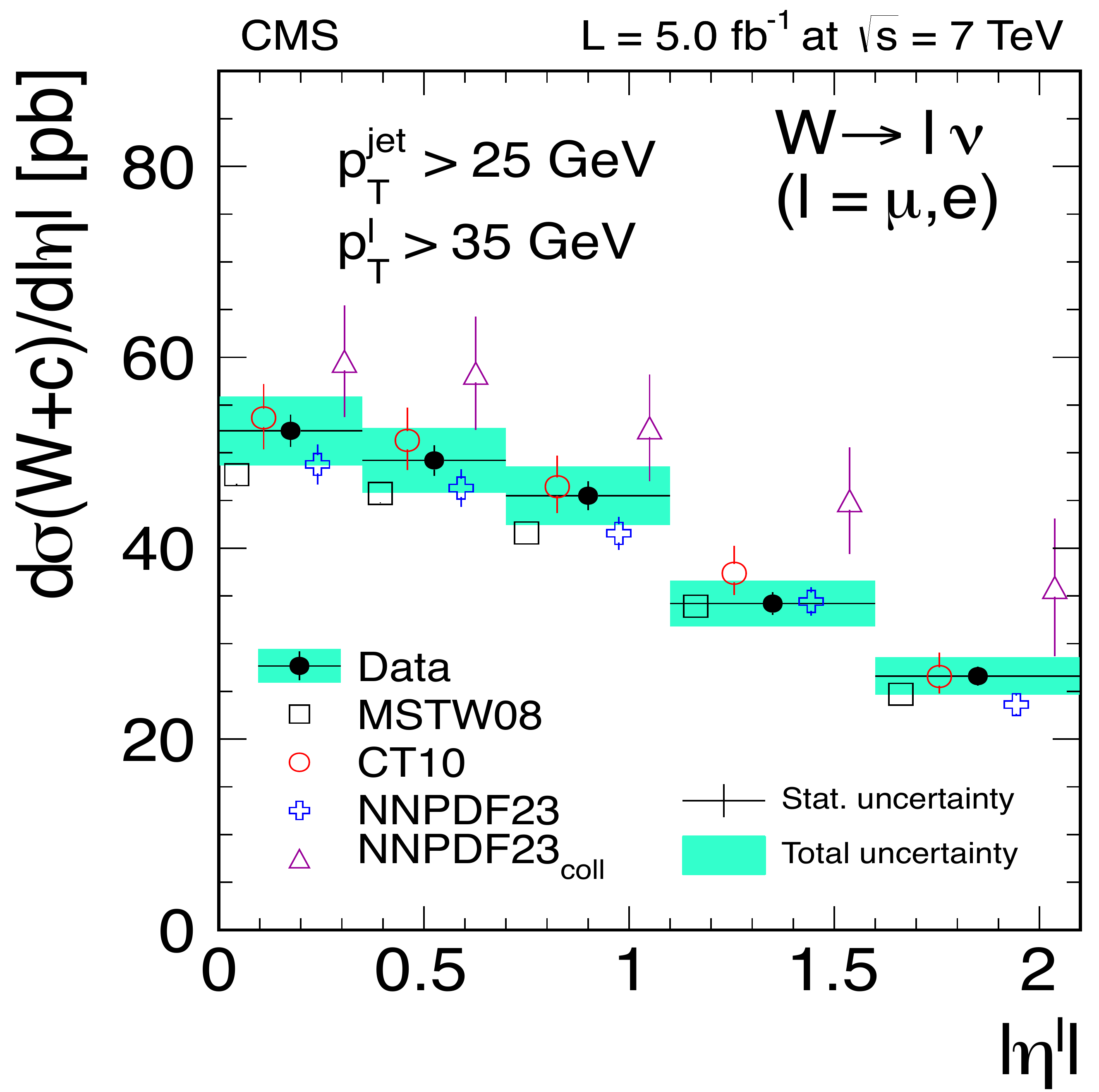}
\end{minipage}
\caption
{The lepton pseudorapidity distribution for the  $W$+charm jet cross section from ATLAS (left) and the $W$+charm 
cross section at CMS (right).}
\label{fig:5}
\end{figure}
}

\newcommand{\figsix}{
\begin{figure}[tph]
\begin{minipage}{0.47\textwidth}
\includegraphics[width=\textwidth,height=0.75\linewidth]{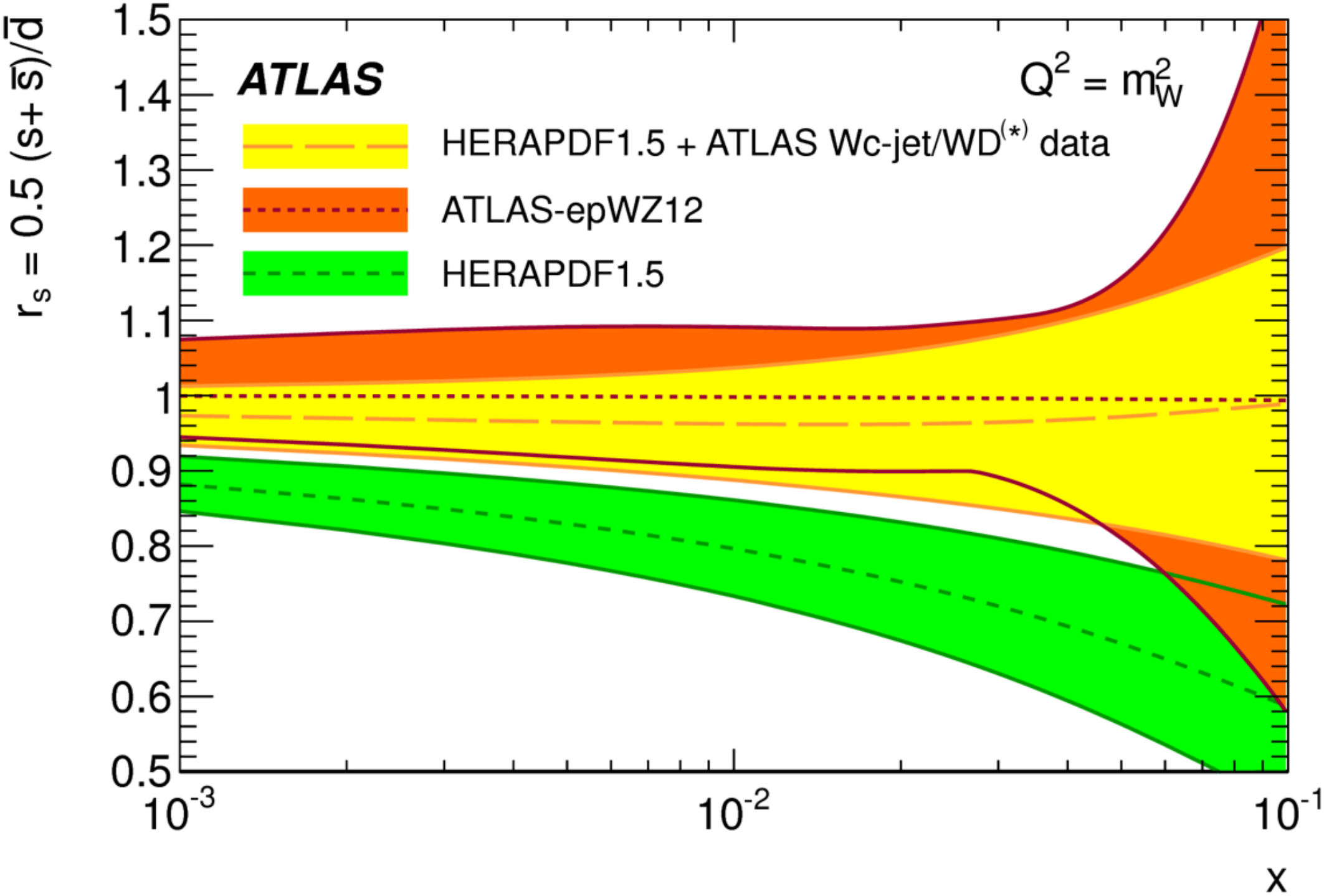}
\end{minipage}
\hspace{0.01\textwidth}
\begin{minipage}{0.52\textwidth}
\includegraphics[width=\textwidth,height=0.8\linewidth]{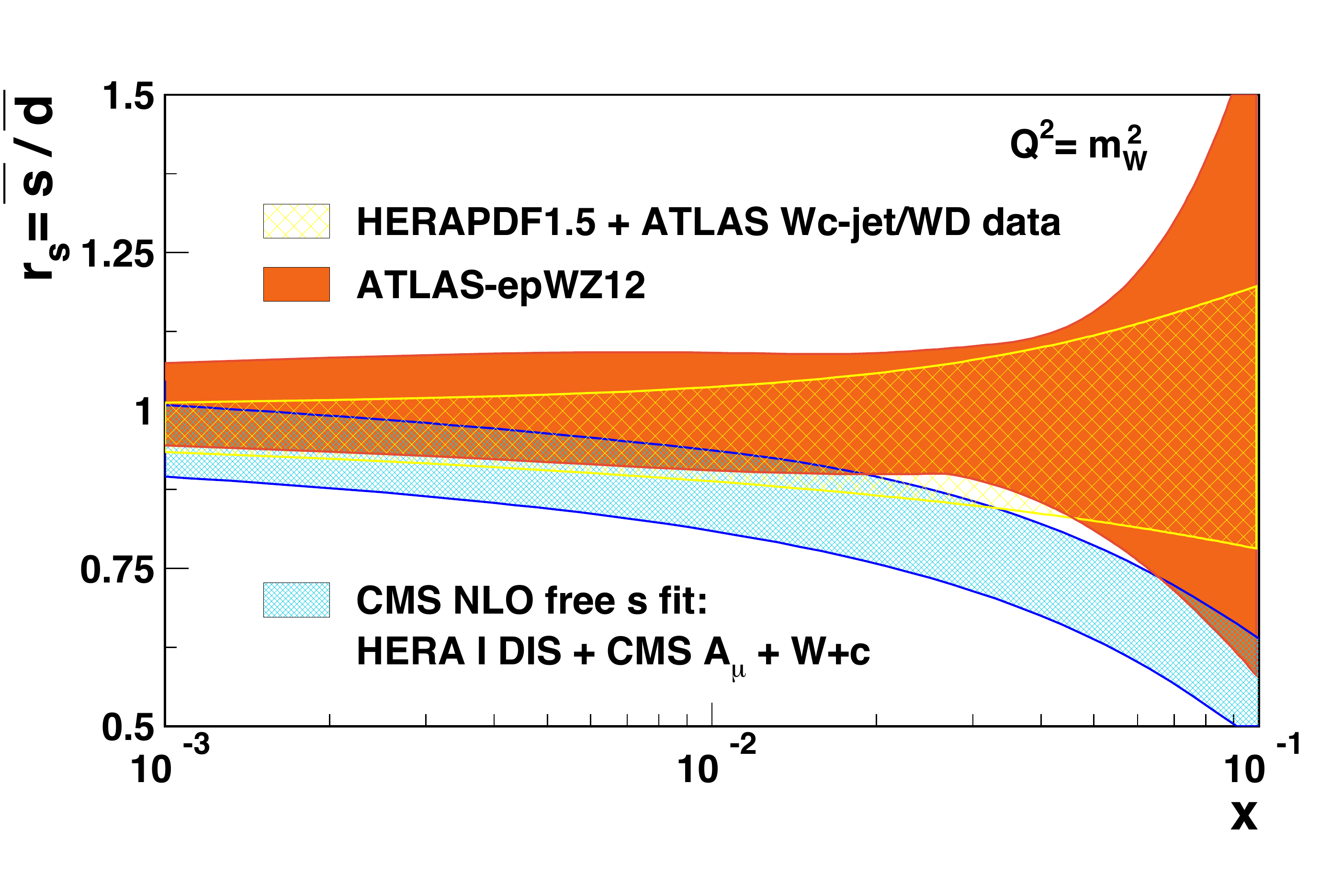}
\end{minipage}
\caption
{The resulting $r_S$ parameter from the ATLAS NNLO, and eigenvector fits (left) and 
both the ATLAS fits and the full CMS NLO fit to the $W$ asymmetry and $W+c$ CMS data (right).}
\label{fig:6}
\end{figure}
}

\section{Introduction}

The LHC~\cite{lhc} is an exemplary machine for the study of perturbative QCD. 
All processes at the LHC take place between quarks and gluons so, given sufficient understanding 
of the hard interaction, any physics process could in principle be used to constrain the parton density 
functions (PDFs) within the proton. Precise knowledge of these PDFs is an essential prerequisite for 
the identification of any possible signature from physics beyond the Standard Model.

During Run 1, the LHC performed extremely well, allowing both the ATLAS~\cite{atlas} 
and CMS~\cite{cms} collaborations to 
collect proton-proton data samples in excess of 25\,fb$^{-1}$ per experiment. 
These large samples are allowing the development of an increasingly large  
and diverse portfolio of precision analyses from each collaboration that are useful for 
constraining the parton distributions in the proton in a kinematic regime beyond any
previously available.

The data presented in this article were collected during 2010 and 2011 and represent only
about one fifth of these Run 1 data sets. 
In addition, the remaining 20\,fb$^{-1}$ of data from the 8\,TeV running in 2012 
available from each collaboration,  are currently being analysed. Results from these 
analyses of the 2012 data will be published in the near future.

When performing fits for the proton parton distributions, different physics processes
provide  information on the different parton initial states. The very precise HERA inclusive 
DIS data typically only tightly constrain the parton distributions at lower-$x$, and are  
sensitive to the gluon at Next-to-leading order (NLO) through scaling violations~\cite{scaling}.
The LHC cross section on the other hand, with two target hadrons,
is sensitive to the gluon distribution and the strong coupling already at leading order (LO) 
for processes including dijet or $t\bar{t}$ production, and to the valence quarks 
at higher $E_{\rm T}$, whereas Electroweak boson production, is sensitive both to the valence 
and sea quark distributions. 
 
For LHC processes, the cross section is generally only calculable after the 
numerical integration over the phase space to cancel the infra-red and collinear divergences, 
so fast convolution techniques are required, such as those implemented by the fastNLO~\cite{fastnlo} 
and APPLgrid~\cite{applgrid} projects.

\section{Jet production and the gluon distribution} 

\figone

Fits to the proton PDFs~\cite{ct10,mstw,abm,herapdf} using only data from experiments with 
lower momentum transfer than 
available at the LHC typically have large uncertainties for the LHC kinematic region, notably 
at high-$x$~\cite{watt}. 
This is most apparent for the gluon distribution, 
where the  difference between the central fits from the different groups
are sometimes  larger than the uncertainties on the individual PDFs themselves. 
This gives rise to a significant uncertainty 
on the predictions of LHC cross sections, such as in Higgs, or $t\bar{t}$ 
production~\cite{watt}. 

For jet production at the LHC, at all but the highest 
$p_{\rm T}$, the cross section is dominated by quark-gluon scattering and so the data 
should provide a significant additional constraint on the gluon distribution with respect 
to that from the HERA or fixed target data alone.

Fig.~\ref{fig:1} shows the new dijet data from both the ATLAS~\cite{atlasjets} and CMS~\cite{cmsjets}
collaborations in different regions of dijet rapidity, spanning a range of 200\,GeV to 5\,TeV 
in dijet mass. The data themselves are reasonably well described over eight orders of 
magnitude in variation of the cross section and a mass range from around 250\,GeV to 5\,TeV
-- somewhat of a triumph for perturbative QCD.

In Fig.\ref{fig:2} can be seen the ratio of 
theory over data for each of the six rapidity ranges seen in Fig.~\ref{fig:1} for the 
ATLAS data. This illustrates that the systematic uncertainties, which are dominated 
by the Jet Energy Scale, are typically between 5-10\% for central rapidities, 
and increase at higher rapidities and higher masses where the data is in any case, 
more statistically limited. In all cases, the experimental uncertainties are 
already comparable to the theoretical uncertainties.
The data are shown compared to the NLO QCD predictions using the  CT10~\cite{ct10}, and HERAPDF fits~\cite{herapdf}, 
and also the atlas epJet13 fit~\cite{epj13} which used the 2.76 and 7\,TeV jet data
from 2010. There appears to be a possible tendency for the CT10 fit to be slightly
harder, with the fit to the ATLAS 2010 data, describing the shape better at high masses.  
The gluon-gluon terms in the full next-to-next to leading order (NNLO) calculation have recently been calculated~\cite{nigel} and 
show approximately a 25\% contribution, although the contribution to the cross section in the 
LHC kinematic region of the gluon-gluon term is small and predominantly at low $E_{\rm T}$.

\figtwo

Quantitative analysis on the level of agreement of the cross section 
with the predictions of the different PDFs~\cite{atlasjets} illustrate 
that  most of the PDFs fits reasonable well, whereas the ABM11~\cite{abm} 
with a softer gluon contribution, is somewhat disfavoured.

\section{QCD analysis}

Both ATLAS and CMS have presented fits to their inclusive jet data --
the ATLAS fit~\cite{epj13} using both the 7 and 2.76\,TeV from 2010 and CMS~\cite{cmsfit} 
using their more recent 2011 inclusive 
data~\cite{cmsincljets}.

Both Collaborations use a similar parameterisation with a more flexible gluon 
distribution. Sum rules are used to constrain many of the 
parameters, and in both the ATLAS and CMS fits the strange distribution is constrained to 
be strictly proportional to the d-type sea, resulting in a 13 parameter fit.
Both collaborations also include the HERA DIS data in their fits, as well as their own data 
which are reproduced using either fastNLO or APPLgrid.

The resulting gluon distribution from the CMS collaboration can be seen in Fig.~\ref{fig:3}.
This also shows the fit resulting from just the HERA DIS data shown as the hatched band.
For the fits from both ATLAS and CMS the gluon distribution is harder than that obtained 
from HERA data alone, with a with a slightly different shape and with a significantly 
reduced uncertainty at higher-$x$.
Most noticeable is the reduction in the uncertainty at high-$x$ at a scale of 10$^4$\,GeV$^2$.  

\figthree

The direct $t\bar{t}$ production cross section is also sensitive to the gluon
distribution and measurements have been performed by both ATLAS~\cite{atlasttbar} 
and CMS~\cite{cmsttbar}. 
Comparisons of the ATLAS data with theory at NLO suggest that the data may be better 
described by the HERAPDF rather than the CT10 PDF, which has a slightly harder gluon 
distribution from the inclusion of the Tevatron jet data.

\section{The strong coupling}

The top production data are also sensitive to the strong coupling, or conversely, to 
the mass top quark itself. By constraining the top mass to a value, 173.2$\pm1.4$\,GeV, 
the CMS collaboration has extracted the strong coupling using an NNLO fit~\cite{cmsttbar}, 
where the result is 
$0.1151^{+0.0033}_{-0.0032}$,
consistent with the world average.

\figthreea

In addition, CMS has also extracted the strong coupling from several jet measurements.
Fig.~\ref{fig:3a} shows the strong coupling extracted using from these jet data 
for the ratio of 3-to-2 jet production~\cite{3to2},
the 3-jet mass~\cite{3jetmass}, and inclusive jet production~\cite{cmsincljets}.
In addition, the value from the fit using the top data is also shown. 
The running of the strong coupling can clearly been seen up to scales nearly 
an order of magnitude higher than measured previously at either HERA~\cite{h1alphas1,h1alphas2,zeusalphas} or the 
Tevatron~\cite{cdfalphas,d0alphas1,d0alphas2}.%

The value of the strong coupling measured using the CMS inclusive jet data~\cite{cmsfit} is
$0.1185^{+0.0065}_{-0.0041}$
again consistent with the world average.

\section{Photon production}

A measuring that has the potential to sample the hard subprocess directly is that of 
prompt photon production. As in the case of jet production, the dominant production mechanism
is quark-gluon scattering. This process has the potential to also constrain
the gluon distribution, although in this case it is also more sensitive to the contribution from 
$u$-$g$ scattering because of the larger charge on the $u$-quark.
A sensitivity study from ATLAS~\cite{photon} using the data on inclusive direct 
photon production~\cite{photondata}
suggests that the 
softer gluon distribution from the ABM11 PDF  is able to describe the 
shape of the cross-section better, but with a lower normalisation. 
Taking into consideration the systematic uncertainties, all the PDFs fit the 
data reasonably well, with the harder gluon distribution from CT10 being less favoured.

\figfour

\section{Heavy Electroweak boson production}

With the production of heavy Electroweak bosons, it is possible to better constrain the 
valence and sea quark distributions. 
Data from the ATLAS~\cite{atlasZ} and CMS~\cite{cmsZ} collaborations on Drell-Yan production are 
becoming rather precise, particularly in the region of the $Z$ resonance, and suggest that 
the NNLO cross section with a number PDFS fitted at NNLO tend to lie somewhat below the 
data for central rapidities.

The charge asymmetry of the  
$W^+$ and $W^-$ data when taken in combination has the potential to largely cancel the 
contribution from the gluon for which the uncertainty can be large.
If the approximate 
equivalence  of quark and anti-quark sea can be assumed then 
much of the sea quark contribution will also vanish, allowing these processed
to provide valuable information on the $u$ and $d$-valence quark distributions.
An additional benefit of taking the charge asymmetry  is the the potential 
to also cancel many of the correlated experimental systematic uncertainties. 

In practice the $W$ kinematics cannot be measured, but fortunately the lepton 
asymmetry remains sensitive to the valence quark distributions.  

The charge asymmetry data from CMS~\cite{cmsW} is shown in Fig.~\ref{fig:4} compared to 
calculations at NLO. The theoretical uncertainties are typically larger than the 
experimental uncertainties at NLO. 
Some discrimination between the data is already visible - particularly for the
default MSTW2008 fit~\cite{mstw}, which lies below the data for small muon 
pseudorapidities. 
This is somewhat improved with the updated MSTW2008 fit including data from 
deuteron scattering,
but still predicts a slightly smaller asymmetry. Note that the HERPDF predicts too 
small an asymmetry at larger rapidities.
Although the $W$ cross section is know to NNLO, differences between the PDFs are still apparent.

\section{Heavy Electroweak boson production with charm}

Events with a $W$ candidate and a charm quark in the final state are directly 
sensitive to the strange quark density.
Recent results from the ATLAS~\cite{atlaswc} and CMS~\cite{cmswc} collaborations 
measure the differential cross section for measuring a $W$ in conjunction with either a 
fully reconstructed charmed hadron, or by identifying a charmed jet by the 
presence of a soft lepton within a jet itself.

\figfive

The differential cross section for the rapidity of leptons from the $W$ decay 
for  charmed jet events from both ATLAS and CMS is shown in Fig.~\ref{fig:5}.
Here the general trend of the predictions suggests that the NNPDF2.3coll
fit~\cite{nnpdf}  predicts the highest overall cross section 
and MSTW2008 the lowest, although the level of agreement with these 
predictions is different between the two measurements.
The ATLAS and CMS measurments are not strictly comparable -- for the  ATLAS measurement 
the  lepton selection is $p_{\rm T}^{\rm lepton}>40$\,GeV whereas for the CMS measurement 
it is is $p_{\rm T}^{\rm lepton}>35$\,GeV --  
although there still may be a tendency for the ATLAS cross section to 
be lower than that expected by the CMS cross section. However, 
is should be noted that in both cases, the treatment of the  charm-to-jet 
fragmentation is treated differently, with CMS extrapolating back the the 
parton level.

\subsection{QCD fits including heavy electroweak boson production}

When including the electroweak  boson data in a QCD fit, both collaborations 
allow the parameters for the strange quark density to vary independently, producing 
a 15 parameters fit, in contrast to the 13 parameter fit with a constrained strange quark 
density used for the fits to the jet data. 

For the Electroweak boson cross sections, APPLgrid is used interfaced to MCFM~\cite{mcfm},
and using an NNLO K-factor obtained from FEWZ~\cite{fewz} in the case of the ATLAS fit to 2010 
inclusive $Z$ and $W$ asymmetry data~\cite{atlasepwz}, and at NLO  
and for the CMS fit~\cite{cmsW}, this time to the 2011 $W$ asymmetry data together the  $W+c$,
with the $W+c$ data extrapolated back to the total charm jet cross section at parton level.

In addition, the new ATLAS measurement includes an {\em eigenvector} fit 
to the 2011 ATLAS $W+c$ data using the HERAPDF1.5 eigenvector set. Here, the data is fit at true hadron level
by fitting  a linear combination of the HERAPDF eigenvector sets with the constraint 
on the strangeness suppression factor released and allowed to vary in the fit.

The distribution of the strangeness ratio, $r_s=(s+\bar{s})/2\bar{d}$ from the ATLAS and CMS fits
can be seen in Fig.~\ref{fig:6}. The left plot shows the comparison of the ATLAS NNLO fit, with
the free strangeness distribution, compared to the HERAPDF1.5 fit, and the ALTAS NLO eigenvector 
fit to the $W+c$. In both the ATLAS fits, a value of $r_s$ close to 1 is obtained. 
On the right, the ATLAS fits are compared to the value obtained from the 
NLO QCD fit to the combined $W$ asymmetry, and $W+c$ data from CMS
The CMS fit also suggests an enhanced strange contribution 
with respect to the HERPDF at low-$x$, although not as large as that predicted 
by the ATLAS fit, being somewhat lower at higher $x$.

\figsix

These data on $W$ production in conjunction with a charm quark
from both ATLAS and CMS are quite recent and potentially extremely 
promising and tools are still being developed to more properly 
include them in a QCD fit, which should shed more light on  
the strange quark density.  

\section{Outlook}

Both ATLAS and CMS have a large, and growing portfolio of precision measurements 
that have the potential to significantly constrain the parton distributions in the proton, 
a small selection of which have been discussed here. 
Higher luminosity data is already available and are being analysed with a view to 
reducing both the statistical and systematic uncertainties of the measurements.

For some of the data, the potential is limited only by the theoretical uncertainty, 
or the available fast convolution grid technology from fastNLO or APPLgrid. 
For some processes, such as jet production, and the $W$+charm production,
calculations are available only at NLO. In these cases, theoretical 
uncertainties are often comparable to, or larger, than those from the data.
However, these new precise data are only now becoming available, and developments 
in both the theoretical calculations and in the fast grid technologies over the last 
few years mean that much of this can, in principle, 
now be used in QCD fits - something that would not have been possible even a few 
years ago. 

The journey towards better understanding of the parton distributions within the proton 
using the LHC data has only just begun, but the significant promise of the new data 
mean that it will be a very interesting time ahead.

\section{Acknowledgements}
The author would like to thank the University of Sussex for funding to attend the 
workshop and the workshop Organising Committee for a most interesting workshop.


\end{document}